\documentclass[10pt, journal]{IEEEtran}
\usepackage{blindtext, graphicx,booktabs,verbatim,colortbl,mathtools,amssymb,commath,url,multirow,color}
\usepackage[ruled,vlined]{algorithm2e}
\SetAlFnt{\small}
\usepackage[table]{xcolor}
\usepackage{verbatim} 
\usepackage{colortbl}
\usepackage{placeins}
\newcommand{\RNum}[1]{\uppercase\expandafter{\romannumeral #1\relax}}
\usepackage[none]{hyphenat}
\usepackage{amssymb,url}
\usepackage{float}
\usepackage{svg}
\definecolor{LightCyan}{rgb}{0.88,1,1}
\graphicspath{{images/}}

\usepackage[numbers,sort&compress]{natbib}
\usepackage[utf8]{inputenc}
\makeatletter
\newcommand*{\rom}[1]{\expandafter\@slowromancap\romannumeral #1@}
\makeatother

\makeatletter
\def\BState{\State\hskip-\ALG@thistlm}
\makeatother

\ifCLASSINFOpdf
\else
\fi

\begin{document}
\newpage
\thispagestyle{empty}

\noindent\begin{minipage}{\textwidth}
    {\Huge\textbf{IEEE Copyright Notice}} \\ 

    \vspace{2cm}
    \Large{\copyright 2020 IEEE. Personal use of this material is permitted. Permission from IEEE must be obtained for all other uses, in any current or future media, including reprinting/republishing this material for advertising or promotional purposes, creating new collective works, for resale or redistribution to servers or lists, or reuse of any copyrighted component of this work in other works.}
\end{minipage}

\vspace{2cm} 

\noindent\begin{minipage}{\textwidth}
    
    \LARGE{Accepted for publication in: \textbf{IEEE Journal on Selected Areas in Communications} \\ \\
    DOI: \textbf{10.1109/JSAC.2020.2986692}}
    
\end{minipage}

\newpage

%
\title{Market Driven Multi-domain Network Service Orchestration in 5G Networks}

\title{\textcolor{red}{\small{This paper has been accepted for publication in IEEE Journal on Selected Areas in Communications (Volume: 38, Issue: 7, July 2020). \\DOI: 10.1109/JSAC.2020.2986692}} \\ [2ex] Market Driven Multi-domain Network Service Orchestration in 5G Networks}

\author{
    \IEEEauthorblockN{Mouhamad Dieye\IEEEauthorrefmark{1}, Wael Jaafar\IEEEauthorrefmark{2}, Halima Elbiaze\IEEEauthorrefmark{1}, Roch Glitho\IEEEauthorrefmark{3}\IEEEauthorrefmark{4}}\\
    \IEEEauthorblockA{\IEEEauthorrefmark{1}Universit\'e du Qu\'ebec \`A Montr\'eal, Montreal, Quebec, Canada \\}
    \IEEEauthorblockA{\IEEEauthorrefmark{2}Carleton University, Ottawa, Ontario, Canada \\}
    \IEEEauthorblockA{\IEEEauthorrefmark{3}Concordia University, Montreal, Quebec, Canada \\}
    \IEEEauthorblockA{\IEEEauthorrefmark{4}University of Western Cape, Cape Town 7535, South Africa \\}

\thanks{
Mouhamad Dieye and Prof. Halima Elbiaze are with the Department of Computer Science, University of Quebec in Montreal, QC, Canada, e-mails: dieye.mouhamad@courrier.uqam.ca, elbiaze.halima@uqam.ca. 

Dr. Wael Jaafar is with the Department of Systems and Computer Engineering, Carleton University, Ottawa,
ON, Canada e-mail: waeljaafar@sce.carleton.ca. 

Prof. Roch H. Glitho is with the Concordia Institute of Information Systems Engineering, Concordia University, Montreal, QC, Canada, and also with the Computer Science Program, University of Western Cape, Cape Town 7535, South Africa, e-mail: glitho@ciise.concordia.ca.}
}

\maketitle



\begin{abstract}
The advent of a new breed of enhanced multimedia services has put network operators into a position where they must support innovative services while ensuring both end-to-end Quality of Service requirements and profitability. Recently, Network Function Virtualization (NFV) has been touted as a cost-effective underlying technology in 5G networks to efficiently provision novel services. These NFV-based services have been increasingly associated with multi-domain networks. However, several orchestration issues, linked to cross-domain interactions and emphasized by the heterogeneity of underlying technologies and administrative authorities, present an important challenge. In this paper, we tackle the cross-domain interaction issue by proposing an intelligent and profitable auction-based approach to allow inter-domains resource allocation. 
 
\end{abstract}


\IEEEpeerreviewmaketitle


\section{Introduction}
Recent years have seen the emergence of services such as tactile internet, multi-player gaming, etc. characterized by their innovativeness and stringent quality of service (QoS) requirements, such as
ultra low latency \cite{Guerzoni2017, Osseiran2014}. 
5G networks have been envisioned to provide flexible and cost-effective service provisioning through several enabling technologies, while ensuring profitability \cite{Akyildiz2016,Ngmn5g2015}. One of the major goals of 5G networks consist of providing ultra reliable end-to-end service delivery through satisfaction of end-to-end QoS requirements \cite{FgIMT2020}. In this regard, Network Function Virtualization (NFV) has been advocated as an effective service provisioning model for network operators to efficiently create, deploy and manage services, thereby allowing efficient and flexible utilization of their limited infrastructures while decreasing capital and operational expenditures. 
Through NFV, a network service (NS) is decomposed into a set of chained virtual network functions (VNFs) running atop of a virtualized infrastructure called Network Function Virtualization Infrastructure (NFVI). In order to model and deploy a service, an efficient approach is required to determine the composition/decomposition of the service, and how to automate selection and control (e.g. VNF creation, placement, migration, monitoring, etc.) of underlying physical or virtual resources and services with certain objectives (e.g. QoS requirements, costs, etc.). This process is referred to as network service orchestration (NSO) and provisioning \cite{Mijumbi2016, deSousa2018}.

Recently, end-to-end service provisioning has been increasingly associated with multi-domain networks where the resources hosting VNFs are owned and controlled by multiple independent network operators \textcolor{black}{located} in different geographical locations \cite{Guerzoni2017, Rosa2015}.
Nonetheless, guaranteeing an end-to-end service delivery with stringent QoS requirements is a challenging task due to several issues related to cross-domain interactions, such as the heterogeneity of underlying 5G technologies including fog and edge computing platforms, various radio access technologies, and different transport and core networks. In addition, as the network infrastructure can be owned and managed by different administrative entities, it also requires taking into account different business models and orchestration approaches \cite{ Katsalis2016, Rosa2015, deSousa2018}. Indeed, a disruption in one resource/domain can carry a detrimental effect to the overall satisfaction of end-to-end QoS requirements. Yet, as of today, no clear consensus on multi-domain service provisioning has been achieved. Accordingly, several studies have noted the lack of support of service provisioning by the ETSI MANO, in the context of multi-domain networks architecture  \cite{Katsalis2016,deSousa2018}.

Thus, new profitable cooperation and effective service orchestration mechanisms are urgently needed to leverage the resources offered by operators to support end-users' QoS and ensure end-to-end service provisioning. Furthermore, as end-user demands and networks themselves (e.g nodes, routes, etc.) are characterized by high dynamism and heterogeneity, future NSO mechanisms should autonomously adapt to various network environments, topologies, and sizes. Finally, given the market competition from third-party operators, e.g. virtual operators and service providers, operators' profitability has to be taken into account in the resource orchestration process.

Albeit mostly in single domain networks, several approaches 
have been explored in the literature in the realm of service provisioning related problems, e.g. resource allocation, slicing, scheduling, etc. However, a number of challenges remain to be effectively tackled in the context of large scale 5G multi-domain networks where emphasis is put on automated network resource sharing, multi-tenancy and cooperative resource provisioning.

In this paper, we focus on the multi-domain network provisioning problem, by taking into consideration automated resource sharing and cooperative resource provisioning. For that, we propose an auction-based NFV orchestration method, where inter-operator interactions and exchange of network resources for service orchestration is translated into buyer/seller transactions and market dynamics \cite{Habiba2018, Zhang2013}. 
Indeed, auctions as a resource allocation and orchestration mechanism have the advantage of being economically efficient through automatic discovery of service chain market value and assignment of limited resources to the bidders who value them the most. Furthermore, it provides compelling opportunities to modern telco actors, namely Infrastructure Providers/Mobile Network Operators (InPs/MNOs) and Service Providers/Mobile Virtual Network Operators (SPs/MVNOs). Briefly, InPs own physical/virtual infrastructure and resources, that are leased to different SPs. The SPs are operators who provide 5G services and yet do not have the required physical infrastructure and resources to meet the demands and requirements of their subscribers. Thus, they rent resources from InPs and, if unused, can lease those resources to other SPs. This business model allows SPs to acquire resources to satisfy their end-users QoS requirements and enables them to reduce new service deployment costs, while InPs increase their profits by leasing their unused network resources within a marketplace. Finally, given a multi-InP marketplace, interested SPs can switch between InPs when market dynamics are not favourable \cite{Habiba2018, Palattella2016, Zhu2016, Jiang2017}. 
 
In practice, self-management through an efficient auction-based framework for large scale and dynamic networks remains a challenging task both from the auctioneer's and bidders' perspective respectively. This is due to the increasing complexity, resulting from the exponential bid space growth, e.g. increased/decreased competition, inflation, new market entry, etc., as in-market resources and market dynamics fluctuate over time \cite{Brero2018, Ausubel2017}. Coincidentally, the advent in recent years of deep reinforcement learning (DRL) algorithms has enabled solving decision-making problems, previously considered intractable due to high-dimensional state and action spaces. In short, DRL algorithms can be exploited to produce fully autonomous agents (SPs) capable of interacting with their environment (market conditions) to learn optimal behaviours (bidding strategies and resource requests), improving over time through trial and error.

The main contributions of this paper are summarized as follows:

\begin{enumerate}
    \item First, we emphasize the importance of multi-domain interactions for end-to-end service provisioning, through the Massive Multiplayer Online Gaming (MMOG) use case. 
    \item Then, we model a market-driven multi-domain interaction framework for resource allocation between an InP and several candidate SPs. We formulate the associated resource allocation problem, where the InP and SPs aim at maximizing their profits by selling/buying in-market resources. We define different marketplace environments, where SPs can compete and/or cooperate in the bidding process, and the InP can decide whether to apply or not fairness among SPs.   
    \item Due to the complexity of the system with several SPs, we propose a distributed multi-agent deep reinforcement learning approach, where we equip each SP with a learning agent able to perceive the environment and takes strategic actions to win auctions, hence satisfies its QoS requirements and increases its profit. Moreover, agents of different SPs may exchange information in cooperation scenarios to improve their mutual profits. 
    \item Through experimental setups, we trained SPs' agents and evaluated the performances of InP and SPs in different marketplace scenarios. It has been shown that a fully competitive market (no cooperation) is the most profitable for InP. In contrast, a cooperative market is in average the most lucrative for SPs. Moreover, double-deep-Q-learning agents outperform other learning-based and learning-free agents in terms of SPs' profits. Finally, impact of some parameters are emphasized.  
\end{enumerate}

The rest of the paper is organized as follows. Section II presents a motivating use case, highlighting multi-domain interactions. The following Section III details the related work. Section IV provides a system model of a market-driven multi-domain interaction framework. Then, the associated problem is formulated in Section V. Section VI describes the investigated auctions scenarios, while section VII presents our proposed solutions. In Section VIII, experimental results are presented and discussed. Finally, Section IX closes the paper.

\section{Massive Multiplayer Online Game use case}
\subsection{Overview}
Massive Multiplayer Online Gaming (MMOG) has increasingly gained popularity to become one of the most lucrative industries with millions of subscribers worldwide with an estimated 55\% of Internet users today to be also online gamers \cite{Lee2010}. MMOG is described as a gaming environment where a large number of active concurrent players 
gather around a shared virtual environment. It is also characterized by its real time requirements to ensure an immersive game-play experience. Unfortunately, as the number of connected subscribers increases, the resource load generated by a game server may induce a degradation of game-play experience 
making the game unplayable to players who eventually quit \cite{Prodan2016}. Typically, to ensure the QoS requirements of its widely distributed subscribers at all times, MMOG operators maintain a rigid multi-server distributed infrastructure with -often- over-provisioned computational and network capabilities. Such design increases operational costs due to potential resource under-utilization and/or capacity shortages caused by sudden peaks in demands \cite{Nae2011}. Alternatively, cloud based MMOG has been increasingly advocated to solve some of these issues \cite{Prodan2016}. 
Also, in order to serve several concurrent players into a unique game session, a current practice is to \textit{parallelize} the game server code and distribute the load across multiple resources, through techniques such as \textit{zoning}, in which the game world is geographically partitioned into disjoint zones that can be assigned to different autonomous computing resources \cite{Prodan2016}.

\begin{figure*}[t]
    \centering
    \includegraphics[scale=0.5]{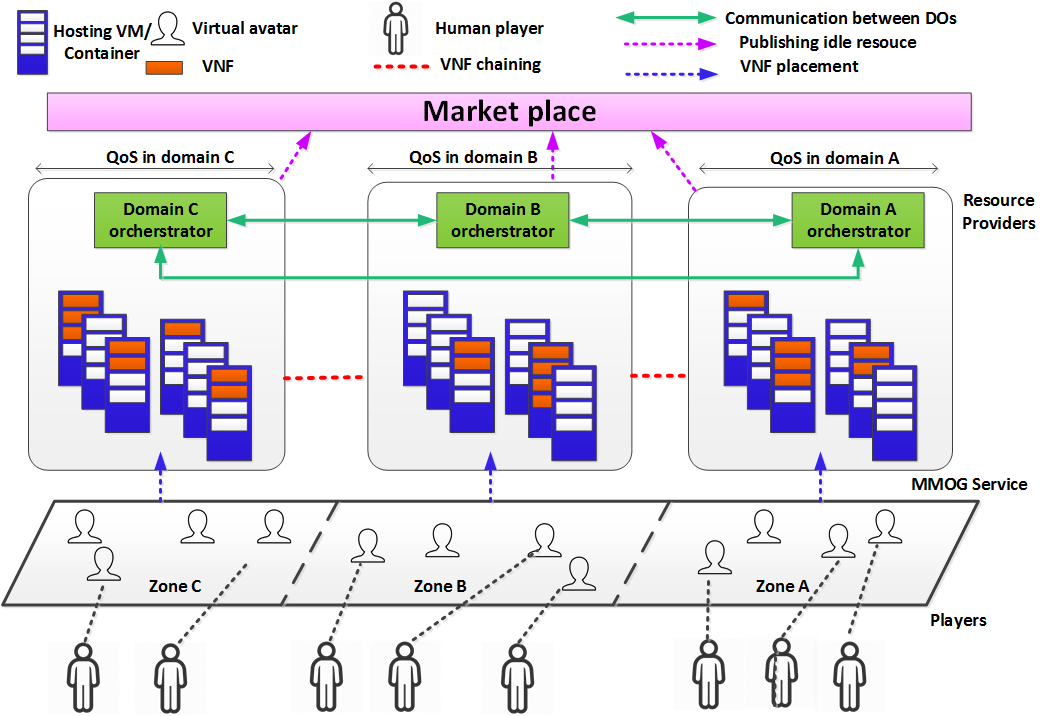}
    \caption{MMOG scenario.}
    \label{fig:MMOG_Model}
\end{figure*}

Similar to \cite{Prodan2016}, we consider an ecosystem consisting of players connected through game providers to game operators who ensure autonomous execution of MMOG sessions by provisioning physical/virtual resources from the worldwide distributed resource providers. Resource providers host the game servers and may serve several independent game operators simultaneously. In this ecosystem, as an incentive to respect the QoS requirements, penalization is due for any SLA (Service Level Agreement) violation.

Further consider the scenario of Fig. \ref{fig:MMOG_Model}, where a game operator is in charge of the zoning MMOG slice to which end-users are connected. The game operator is tasked with ensuring QoS requirements regardless of users' demands variation and locations. To this end, it can strategically deploy a game server service as a VNF chain, with each VNF handling a specific game-play task such as command emulation, character management, game physics/logic responses, graphics rendering, etc. It is worth noting that the VNF chain composition may vary, depending on the capabilities of devices (mobile, desktop, etc.) used to play the game. Indeed, some end-users may not have, for instance, the required codecs to output the gameplay, thus requiring transcoding. Also, other end-users, with limited Internet bandwidth, wouldn't obtain full resolution graphics rendering.
Hence, in order to provide the service with regards to specific QoS requirements, the game operator must strategically embed VNFs by leveraging the virtual/physical resources offered by independent resource operators.  

\subsection{Requirements}
Achieving optimal placement of VNFs remains a challenging issue in the MMOG scenario due to the inherent goals associated. First, the game play service must be delivered to the end-users with strict end-to-end QoS requirements, notably end-to-end low latency, as it constitutes the leading reason why end-users drop a game. Thus end-to-end latency is vital from the profitability perspective \cite{Prodan2016}. Second, the chosen service provisioning strategy must be adaptable to both small and large scale scenarios, as the number of players involved is large and time-varying. Third, game operators are to be considered with limited budgets and an emphasis is put on profit maximization, e.g reducing the amount of penalties to pay for SLA violations.

In our scenario, a possible way to deploy a game server service by the game operator, within a single domain, is to leverage a single resource operator to host all required VNFs. However, given the multi-dimensional nature of QoS requirements and the time-varying users' behaviours, it is very hard to provision and ensure timely service delivery, without over-provisioning resources \cite{Buyya2010}. Worse, as there are intermediary networks between the end-users and the resource operators hosting a game server service, ensuring the QoS requirements within a single domain does not guarantee satisfaction of end-to-end QoS requirements.

Alternatively, a game operator could deploy its game server service by efficiently and seamlessly leveraging offered resources by multiple resource operators. Hence, subscribers' heterogeneous QoS requirements can be cooperatively satisfied. Moreover, embedding VNFs closer to customers, as in an edge/fog computing design, would achieve low latency services and simplify delegated monitoring and maintenance tasks \cite{Rosa2015}. 
Also, as each type of domains has its own strengths and weaknesses in terms of network characteristics, e.g. bandwidth, latency, and computing power, embedding VNFs across multiple domains may enhance service performances and cost efficiency. In sum, a game operator can chain VNFs from different operators into a single functional service that needs to operate over heterogeneous network infrastructures owned by different providers. However, a fundamental step to this is to provide a profitable and viable cooperation mechanism to provision and manage VNFs (and services), thereby enabling multi-domain VNF chaining. In the proposed scenario, we assume resource operators to be available through a market where they publish their resources. These resources can be leveraged by game operators to compose their VNF chains and services. Obviously, the proposed service provisioning mechanism requires not only to be efficient in large scale environments, but also to adapt to varying network and market conditions, such as available resources, workload variation, increased competition, price inflation, etc.

\section{Related Work}

Network operators rely on paradigms such as Software-Defined Networks (SDN) and NFV to efficiently create, deploy and manage services to serve their subscribers. By decoupling the physical and logical network layers, an operator is able to model a service from end-to-end by abstracting and automating the control of physical and virtual resources. This orchestration process comes down to an automatic selection and control of multiple resources, services and controllers, in order to satisfy predefined objectives \cite{deSousa2018}.

\subsection{SDN-based Orchestration}
Leveraging SDN technology to separate the control and data plane has brought to light several challenges especially for multi-domain systems. A typically important issue consists of determining the placement of SDN controllers in order to achieve optimized control over physical/virtual resources. In this matter, several papers have investigated SDN controller placement in a single-domain network considering various objectives, such as minimizing resource utilization, overall operational costs or network delay \cite{Bari2013,Lange2015,Wang2017}.  
For multi-domain networks, \cite{Katsalis2017} investigated SDN implementations over converged wireless-optical networks and proposed abstractions and virtualization techniques to integrate virtual wireless and optical resources in a framework called CONTENT. Authors in \cite{Wang2016} proposed a scalable SDN architecture for multi-domain and multi-vendor networks by implementing a coordinator controller to enable cooperation among different SDN administrative domains. In \cite{Figueira2015}, the authors summarized the challenges of SDN multi-domain orchestration and control before proposing a hierarchical SDN framework that aims at simplifying network control and orchestration.

\subsection{NFV-based Orchestration}
From the NFV perspective, numerous research studies (\cite{Mijumbi2016, Bhamare2016} and references therein) investigated service provisioning through optimizing the VNF chain instances placement over virtual/physical resources of a network \cite{Mijumbi2016, deSousa2018, Abu_tnsm2017}. 
Whereas, ETSI has concentrated its efforts in the standardization of NFV management and orchestration. Indeed, ETSI set the standard's requirements, specifications and architectural framework, called NFV-MANO (NFV Management and Orchestration) \cite{nfvArchitecture2014}.
Its role, mainly through the NFVO (NFV Orchestrator) and the VNFM (VNF Manager), is to enable VNF operations, e.g. orchestration, lifecycle management, 
across computing resources within a single administrative network domain.  

Besides VNF placement and resources orchestration, the management of lifecycle operations, such as creation, monitoring, release, and dependencies, and the interactions with other components, e.g. marketplaces to acquire new resources and OSS/BSS (Operation Support System/Business Support Systems), are also important to satisfy end-to-end QoS and realize profitability. Hence, within the NFV-MANO framework and in the context of multi-domain NFV systems, the number and locations of NFVO and VNFM functional blocks are critical to the system's overall scalability and performances. For instance, given that in order to properly satisfy end-to-end latency requirements, latencies between VNFs, Element Managers (EMs), Virtualized Infrastructure Managers (VIMs), NFVOs and cooperating VNFMs, must all be taken into account. Apart from \cite{abuLebdeh2018, Abu_tnsm2017}, very few have investigated the MANO functional blocks placement problem.  

\subsubsection{NFV Single-Domain vs. Multi-Domain Orchestration}

In order to provision and orchestrate network services with end-to-end QoS requirements, a service operator may consider the use of physical/virtual resources and/or services of other operators. The orchestration process 
being noticeably different between a single-domain and multi-domain networks. Indeed, a domain orchestrator has only control over resources within the operator's administrative boundaries \cite{deSousa2018}. Its role covers managing network services' lifecycle (by interacting with other components to control VNFs) and associated service provisioning resources (e.g. computing, storage, communication, etc.).  
Typically, a single-domain orchestrator oversees and controls all resources and services within its domain, by leveraging the ETSI NFV-MANO framework. In contrast, orchestration in multi-domain networks is more difficult, given the incomplete knowledge and control of resources offered by independent providers. From a service composition perspective, multi-domain specification of an end-to-end QoS requirement differs from that in a single-domain network since in the latter a single-domain QoS requirement is provided. Meanwhile, due to the heterogeneity of multi-domain infrastructures and administrative control, partial QoS requirements may be needed for each domain, thus constituting a significantly different set of constraints in comparison to a single domain network orchestration problem \cite{Campbell2013, Abu_tnsm2017}. 

As of today, no standards exist for multi-domain orchestration \cite{deSousa2018}. A number of multi-domain orchestration frameworks have been advocated in the literature, including T-NOVA \cite{Kourtis2017}, SONATA \cite{Draxler2017} and ONAP \cite{Onap}.
Meanwhile, others proposed and investigated NFV orchestration architectures and use-cases inspired from the single-domain ETSI NFV framework. For instance, Rosa et al. \cite{Rosa2015} proposed MD2-NFV where three use case scenarios are studied to highlight the benefits of distributed NFV. In \cite{nfvArchitecture2015, nfvmulti2018}, distributed MANO orchestration is discussed and three models are identified: hierarchical, flat (or peer-to-peer) and hybrid. Most works have adopted hierarchical orchestration in their proposals. ETSI report \cite{nfvArchitecture2018} proposes a two-layer hierarchical architecture addressing end-to-end network services provisioning across two administrative domains, and the adaptation of the NFV-MANO framework to generalized multi-domain networks. In \cite{Sciancalepore2016}, an extension of the ETSI NFV MANO framework to enable joint orchestration of VNFs and edge computing applications is discussed. 
Finally, \cite{Borjigin2018} proposes a scheme named DARA to distribute network resources through a double auction approach. Their design considers three actors: an NFV broker, customers and SPs. The centralized broker collects SPs' resources to supply customers, while maximizing SPs' profits. Unlike previous works, \cite{Borjigin2018} is the first in the literature to propose a multi-domain orchestration scheme. 

In this work, we focus on the issue of NFV-based orchestration for multi-domain networks. We thus propose an auction-based method in which inter-operator interactions and exchange of network resources for service orchestration are encompassed into buyer/seller transactions and market dynamics. Namely, we turn SPs into buyers and InPs into sellers. We further investigate different realistic market conditions to ensure profitability of all actors within the networks. To alleviate the scalability burden of such an auction framework, we leverage the benefits of multi-agent DRL in a heterogeneous and dynamic environment. Hence, SPs and InPs are able to engage into an autonomous negotiation/cooperation mechanism, where profitability and partial QoS requirements' satisfaction are taken into account. 
Besides \cite{Borjigin2018}, this work is among the pioneers in proposing and investigating auction-based NFV multi-domain orchestration from the algorithmic aspect, rather than the architectural aspect. Moreover, to the best of our knowledge, this is the first work to leverage distributed/collaborative intelligence within SPs to achieve autonomous bidding/resource allocation behaviours within resources' marketplace. 

\section{System Model}
For illustrative and simplicity purposes, we hereby describe a 5G mobile network scenario, where two main stakeholders co-exist: (1) an InP as the owner/controller of network resources, including base stations, radio spectrum licences, etc. The latter can be virtualized and leased to (2) SPs, who leverage these resources to offer various innovative and tailored services to attract more users - and profit - to the network \cite{Kiiski2006}. Since resources can be abstracted and sliced into virtual resources, several SPs can then co-exist under the same InP, hence contributing in the InP's expenditure savings. At the same time, by leveraging the logical isolation between these resources, an SP is free to use its allocated resources from the InP to accommodate the heterogeneous provided services to its own subscribers, with respect to QoS and SLA requirements, priority, etc.

\begin{figure}[t]
    \centering
    \includegraphics[scale=0.21]{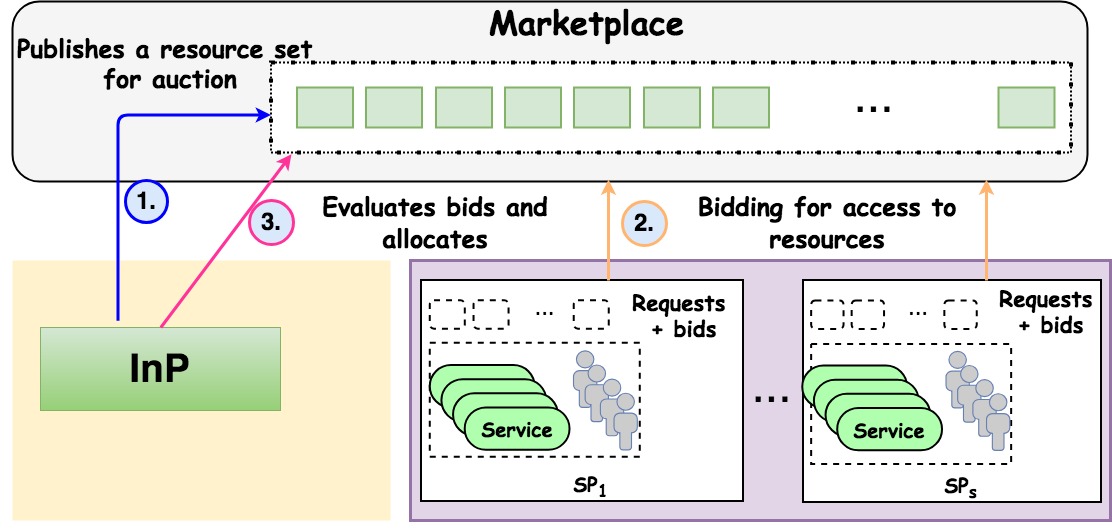}
    \caption{Bidding process.}
    \label{fig:biddingProcess}
\end{figure}

To do so, we consider that the InP publishes its available resources to a free marketplace as shown in Fig \ref{fig:biddingProcess}. Then, the SPs compete for accessing these limited resources through a combinatorial auction process, allowing them to define their bids as combinations of discrete sets of resources, required to satisfy their subscribers needs. Naturally, both the InP and SPs aim at maximizing their own profits. For simplicity, we assume that the InP has no subscribers, and hence does not take part in the bidding process. We assume also that each SP is independently operated, self-interested, and bids using an initially fixed budget. When an SP's budget is exhausted, it leaves the market, and therefore cannot participate in the next bidding rounds. In this context, SPs submit their requests for services with QoS requirements and a bidding offer to the InP. The latter evaluates all requests and bids, then selects the SPs to serve for a given time period. We also incentivize SPs to bid truthfully by assuming a penalty fee to pay each time an SP's bid is rejected, given that it fails to serve its subscribers. 

In order to provide end-to-end service orchestration with stringent QoS requirements, an SP decomposes the service into VNF chains to deploy within network resources managed by InPs. Typically, in a single domain network setting with a centralized NFVO \cite{nfvArchitecture2014}, QoS requirements are generally specified on a higher granularity level. However, in multi-domain networks with distributed and independently managed orchestrators, specifying a global QoS requirement for a given service is inefficient due to to the heterogeneity of underlying networks in terms of available capacity, and to the ubiquitous locations of end-users. Consequently, being inspired from \cite{Duan2010}, we consider a method in which a global QoS requirement ($Q$) is partitioned into partial QoS requirements, i.e. $Q_{InP_i} = \mathcal{P}(Q)$, with $\mathcal{P(\cdot)}$ denoting the partition function. The partial QoS depends on the serving InP's capacity, e.g. available network resources, connectivity, etc. Once the partial QoS requirements are set, the SP must ensure access to the limited resources by competing against other SPs in auctions. To this end, it must be able to derive an optimal bidding strategy regardless of market conditions.

For simplicity, we assume a single InP. Note that our model can be easily extended to multiple InPs, with SPs being able to switch between serving InPs as market conditions become detrimental for them over time, for instance due to sustained inability to satisfy end-users' requests, i.e. loss of profits.

\section{Problem formulation}
Consider a cellular network consisting of one base station (BS) owned and managed by an InP $i$. The InP plays the role of an auctioneer who serves several SPs within $\mathcal{S} = \{1,2,\ldots, |\mathcal{S}|\}$, where $|.|$ is the cardinality operator. Each SP $s$, playing the role of a bidder, is assumed to be in charge of $\mathcal{N}_s = \{1,2,\ldots,|\mathcal{N}_s|\}$ subscribers.

We assume that the available radio spectrum resource is of a total bandwidth $W$, and is divided into $\mathcal{C}=\{1,2, \ldots,|\mathcal{C}|\}$ orthogonal subcarriers. The SPs compete to get these subcarriers.
Also, let $a_{c,s}^t \in \{0,1\}$ be the binary variable indicating whether subcarrier $c \in \mathcal{C}$ has been allocated at the end of time $t$ to SP $s$,  following the bids evaluation by the InP, or not. We assume that for each subcarrier $c$, an undisclosed minimum operational price $\lambda_{c,min}^t$ is estimated and accounted by the InP. Moreover, each SP $s$ has a current bidding budget $\mathcal{B}_{s}^t$, that fluctuates with the bidding rounds where SP $s$ participates.  

Time is considered slotted, where a set of observations time periods are defined as $\mathcal{T}=\{1,2, \ldots |\mathcal{T}|\}$ with $t\in \mathcal{T}$. The end of each time period coincides with the end of the auction and the decision-making process for resources allocation by the InP. Let $y_{s,i}^{t,c} \in \{0,1\}$ be the binary variable indicating if a bid submitted by an SP $s$ is accepted by the InP $i$ at time $t$ or not. 

At the beginning of each time period $t$, SP $s$ sends its request $z_{s}^{t} = \{u_s^t, b_s^t\}$, where, for simplicity, $u_s^t$ is the QoS requirement (e.g. minimum data rate), demanding $c_s^t$ subcarriers to satisfy its $|\bar{\mathcal{N}}_s^t|$ subscribers in time $t$, and $b_s^t$ is the bid that it is willing to pay to access the required resources.

We assume that $b_s^t$ is limited by a maximum bid  $b_{s,max}^t$. Moreover, we define the penalty that SP $s$ has to pay when it fails to satisfy its subscribers' demands as:
\begin{equation}\label{eq:penaltySP}
    \delta_s^t =  |\bar{\mathcal{N}}_s^t| \cdot \delta \quad \forall t \in \mathcal{T},~s \in \mathcal{S},
\end{equation}
where $\delta$ represents a unitary fee. 

From the InP's perspective, 
resource allocation through an economic driven optimization model yields two types of benefits : 
\begin{enumerate}
    \item InP resource-centric: better resources utilization, availability and performances (e.g. data rates).
    \item SP-centric: Increased profits, reduced costs and fairness.
\end{enumerate}
Let $R_i^t$ be the revenue obtained by InP $i$ at time $t$, given by:
\begin{equation}
\textcolor{black}{R_i^t = \sum_{s \in \mathcal{S}} y_{s,i}^{t,c} \cdot b_s^t \quad \forall t \in \mathcal{T},}
\end{equation}
and $R_i^{Total} = \sum_{t \in \mathcal{T}} R_i^t$ is the total revenue garnered by the InP.
The latter is deemed to make a profit at time $t$ when $R_i^t \geq \sum_{c \in \mathcal{C}} \lambda_{c,min}^t$, otherwise it loses money. Hence, the profit maximization problem of the InP can be formulated as follows:
\begin{subequations}
\label{eq:objectiveFunctionInP2}
	\begin{align}
	\max_{a_{c,s}^t} & \quad 
	R_i^{Total} \\
	\text{s.t.}\quad & \sum_{s \in \mathcal{S}} c_s^t \cdot y_{s,i}^{t,c} \leq |\mathcal{C}|\; \quad \forall t \in \mathcal{T}, \label{eq:resourceConstraint} \\
	&\sum_{s \in \mathcal{S}} c_s^t \cdot y_{s,i}^{t,c} \geq \Omega(u^t_s) \quad \; \forall ~t \in \mathcal{T} ,s \in \mathcal{S} \label{eq:resourceConstraint3}\\
	&\sum_{c \in \mathcal{C}} \sum_{s \in \mathcal{S}} a_{c,s}^t \leq 1\quad  \; \forall t \in \mathcal{T},  \label{eq:resourceConstraint2}
	\end{align}
\end{subequations}
where $\Omega(.)$ denotes a function used by the InP to compute the minimum required subcarriers to satisfy a given data rate. Constraints (\ref{eq:resourceConstraint}), (\ref{eq:resourceConstraint3})  and (\ref{eq:resourceConstraint2}) ensure that the InP does not over-allocate its resources, or allocate the same resource to different SPs, while allocating enough resources to satisfy SPs' QoS requirements. 
In order to maintain its level of profitability, the InP must also take the necessary steps to ensure that the value of its resources does not depreciate over time.

\begin{table}[t]
\centering
\caption{Notations}
\label{my-label}
{%
\begin{tabular}{|l|l|}
\hline
$\mathcal{C}$                 & Set of subcarriers\\ \hline
$\mathcal{S}$                 & Set of SPs \\\hline
$\mathcal{T}$                 & Discrete time system  \\ \hline
$\mathcal{N}_s$                 & Set of subscribers for SP $s$\\ \hline
$a_{c,s}^t$                 & 1 if subcarrier c is allocated to SP $s$ at time $t$; 0 otherwise  \\ \hline
$b_s^t $                 & Bid sent by SP $s$ at time $t$  \\ \hline
$b_{s,max}^t$                 & Maximum bid by SP $s$ at time $t$ \\ \hline
$\mathcal{B}_s^t$                 & Budget of SP $s$ at time $t$  \\ \hline
$\bar{\mathcal{N}}_s^t$                 & Set of subscribers of SP $s$ to be served at time $t$ \\ \hline 
$R_i^t$                 &  Revenue obtained by InP $i$ at time $t$ \\ \hline
$R_i^{Total}$                 &  Total revenue garnered by InP $i$ \\ \hline
$R_s^t$                 &  Revenue gained by SP $s$ at time $t$ \\ \hline
$R_s^{Total}$                 &  Total revenue gained by SP $s$ at time $t$ \\ \hline
$u_s^t$                 & Minimum data rate for request sent by SP $s$ at $t$  \\ \hline
$y_{s,i}^{t,c}$                 & 1 if bid submitted by SP $s$ at time $t$ is accepted by InP $i$; 0 otherwise \\ \hline
$z_{s}^{t}$                 & Request sent by SP $s$ at time $t$  \\ \hline
$\lambda^t_{c,min}$                 & Minimum reserve price for subcarrier $c$ at time $t$  \\ \hline
$\delta_s^t$                 & Penalty fee of SP $s$ at time $t$ \\ \hline
\end{tabular}}
\end{table}

From a SP's perspective, the objective is to maximize its profit while satisfying its subscribers' demands. Depending on the market conditions and number of subscribers to serve, a variety of bidding behaviours may be suitable, where the SP evaluates the risks associated to under-market bids. 
We denote by $R_s^{Total}= \sum_{t \in \mathcal{T}}R_s^t$ the total revenue obtained by SP $s$, where $R_s^t$ is the revenue obtained at time $t$, written as:
\begin{equation}\label{eq:SPProfit}
  R_s^t = (y_{s,i}^{t,c} \cdot (b_{s,max}^t - b_s^t )) - \delta_s^t,
  \quad \forall t \in \mathcal{T}, b_s^t \leq b_{s,max}^t,
\end{equation}
The profit problem of SP $s$ is formulated as:
\begin{subequations}
\label{eq:objectiveFunctionInP3}
	\begin{align}
	\max_{b_{s}^t} & \quad 
	R_s^{Total} \\
	\text{s.t.}\quad & b_s^t \leq b_{s,max}^t \leq \mathcal{B}_s^t \quad \forall t \in \mathcal{T}, \label{eq:budgetConstraint} \\
	&\mathcal{B}_{s}^t > 0 \quad \forall s \in \mathcal{S}, \; t \in \mathcal{T}.  \label{eq:budgetConstraint2}
	\end{align}
\end{subequations}
All the defined variables are summarized in Table \ref{my-label}.

A fundamental aspect to note is that auctions differ not only by the associated rules but also by the auction environment. Hence, auctions can be studied in a wide range of environments, with varying numbers of sellers/buyers, number of resources within the marketplace, exchanged information, etc. \cite{Cramton2006}. In our work, we define several scenarios under different auction settings and market conditions. Moreover, since the InP and SP formulated problems generalize the combinatorial allocation/auction problem, in particular the NP-complete winner determination problem \cite{Cramton2006}, heuristic algorithms are needed to find optimal or near-optimal solutions in polynomial times.  


In the next sections, we detail the scenarios to investigate. Then, we expose the developed solutions for the InP and SP formulated problems. 

\section{Auction scenarios}\label{auction_desc}
We present in this section, the auctions scenarios to be considered and evaluated. 
In particular, we consider two situations: \textit{bidding war} and \textit{bidding collusion}. In the former, bidders attempt to outbid each other in their pursuit of network resources. In the short term, this would represent the optimal situation for the InP as it generally leads SPs to bid over the real-value price, thus substantially increasing its profit. Adversely, this may be the worst situation for poor SPs, as it can push them out of bidding rounds and give more power to wealthy SPs to control the market. 
It has been shown in \cite{Maille2009, Bhaskar2002} that oligopolistic/oligopsonic coordination, i.e. \textit{bidding collusion}, is more likely to occur in this situation, aiming at maximizing the collusion-players profits at the expense of other players' and auctioneers' welfare. Bidding collusion in telecommunication networks is a realistic assumption, and has been well-documented in the following works \cite{Maille2009, Massey2008}. The problem is further amplified by technology-aided price fixing algorithms, for which collusion is much harder to detect.
\subsection{Scenario 1}
We first investigate an auction environment where the allocation policy of the InP is known to all bidders, that is the bidders with the highest bids are always selected, for each time period. In this scenario, we assume that SPs may realistically have different initial budget powers, and may pursue resources in a conservative or an aggressive way, depending on their initial budgets and growth in number of subscribers. Moreover, as a result of the InP policy, there is a risk of starvation, hence pushing poor SPs to leave the market while allowing wealthy SPs to expand their control on resources' prices. As a consequence, an oligopsony may be created and the profits of the InP may be threatened \cite{Pla2015}. To avoid this situation, the InP introduces a fairness model where every SP must be served at least once within a fixed number of bidding rounds, denoted $\tau_{th}$. Precisely, the InP keeps track of the number of consecutive wins and losses for each SP, then establishes SPs' priority ranks.  
Finally, we assume that the InP analyzes the interest sparked over time by its resources within the market, i.e appreciation or depreciation, and adjusts its minimum asking price $\lambda_{c,min}^t$ accordingly in order to maximize its revenues or rejuvenate interest for its resources.

\subsection{Scenario 2}
  
Pushing further the first scenario, we analyze the behaviour dynamics of SPs within the marketplace in this second scenario. We assume here that the InP does not apply a fairness mechanism, but rather attempts to detect and punish instances of collusive cooperation among wealthy bidders. In order to allow poor SPs remain competitive, the InP permits competitive cooperation (or coopetition) among them. \textit{Coopetition} is the strategy in which bidders, with partial congruent goals, cooperate and compete simultaneously by sharing partial information \cite{Brandenburger1996,Dagnino2009}.
In our paper, we call by poor, every SP with an initial budget under a predefined threshold.
		     
\section{Proposed solutions}
In this section, we present the proposed strategies to adopt by the InP and SPs aiming at maximizing their profits. 

\subsection{From the InP's Perspective}\label{InpSolutions}
Determining the winners in combinatorial auctions is computationally complex due to the problem's NP-completeness. A synopsis of the problem and main approaches to solve it are presented in \cite{Cramton2006}. For our work, we leverage an algorithm adapted from CABOB (Combinatorial Auction
Branch On Bids) \cite{Sandholm2005}, that is an optimal tree search algorithm based on a mix of linear programming and branch-and-bound techniques \textcolor{black}{(interested readers are referred to \cite{Sandholm2005} for further details)}. This approach is detailed in Algorithm \ref{algo:cabob}.
Its main idea consists of maintaining a bid graph $\mathcal{G}$ where a branch and bound Depth First Search (DFS) method is applied to find the most profitable bids for the InP, i.e. to be selected for service. 

\begin{algorithm}[t]
\SetAlgoLined
 $\lambda_{min} = \lambda^t_{c,min}$\;
 $\mathcal{F}:\{G_1, G_2, ..., G_{|\mathcal{F}|}\} \gets$ DFS(G) \;
 Compute upper bound $\mathcal{U}_f$ \textbf{for} f in $\mathcal{F}$\;
 \If{$\sum_{i=1}^{|\mathcal{F}|} \mathcal{U}_f \leq \lambda_{min}$}{
    return $0$\;
 }
 Compute lower bound $\mathcal{L}_f$ \textbf{for} f in $\mathcal{F}$\;
 \If{$g+\sum_{i=1}^{|\mathcal{F}|} \mathcal{L}_f > f^*$}{
    $f^* \gets g+\sum_{i=1}^{|\mathcal{F}|} \mathcal{L}_f $; ~  $\lambda_{min} \gets \lambda_{min} + g+\sum_{i=1}^{|\mathcal{F}|} \mathcal{L}_f - f*$\;
 }
 \eIf{$|\mathcal{F}| > 1$}{
 
 $\mathcal{F}^* \gets 0$; ~ $\mathcal{U}' \gets \sum_{i=1}^{|\mathcal{F}|} \mathcal{U}_f $; ~ $\mathcal{L}' \gets \sum_{i=1}^{|\mathcal{F}|} \mathcal{L}_f $\;
 
    \For{k in $\mathcal{F}$}{
        \textbf{If} {$\mathcal{F}^* +  \mathcal{U}' \leq \lambda_{min}$ \textbf{then} return $0$\;}
        $g'_f \gets \mathcal{F}^*  + (\mathcal{L}' - \mathcal{L}_f)$; ~ $f^*_{old} \gets f^*$\;
        $f^*_k \gets CABOB^*(G_k, g+g'_k,\lambda_{min}-g'_k)$\;
        $\lambda_{min} \gets \lambda_{min} + (f^* - f^*_{old})$; ~ $\mathcal{F}^* \gets \mathcal{F}^* + f^*_k $ ; ~ $\mathcal{U}' \gets \mathcal{U}' - \mathcal{U}_f$ ; ~ $\mathcal{L}' \gets \mathcal{L}' - \mathcal{L}_f$\;
    }
    return $\mathcal{F}^*$\;
 }{
    $\Theta_s \gets \{ c : a^t_{c,s} = 1 \}$ ; ~ $\Theta_{s'} \gets \{ c : a^t_{c,s'} = 1 \}$\;
    Select next bid set $b^*_s = \{b_s^t \cup \Theta_s\}$ to branch on; ~ $G \gets G - {b^*_s}$\;
    \For{$s' \in \mathcal{S}$}{
        \textbf{If} {$s' \neq s$ and $\Theta_s \cap \Theta_{s'} \neq \emptyset$}
		\textbf{then} $G \gets G - {b^*_{s'}}$\;
    }
    $f^*_{old} \gets f^*$; ~ $f_{in} \gets CABOB^*(G, g+b^t_{s},\lambda_{min} - b^t_{s})$; ~  $\lambda_{min} \gets \lambda_{min} + (f^* - f^*_{old})$\;
    \For{$s' \in \mathcal{S}$}{
        \textbf{If} $s' \neq s$ and $\Theta_s \cap \Theta_{s'} \neq \emptyset$ \textbf{then} $G \gets G \cup {b^*_{s'}}$\;
    }
    $f^*_{old} \gets f^*$\;
    $f_{out} \gets CABOB(G,g,\lambda_{min})$\;
	$\lambda_{min} \gets \lambda_{min} + (f^* - f^*_{old})$ ; ~ $G \gets G - {b^*_{s}}$\;
    return $max(f_{in}, f_{out})$\;
 }
\caption{CABOB*(G, g, $\lambda^t_{c,min}$)}\label{algo:cabob}
\end{algorithm}
Graph nodes represent the SPs' bids that can still be appended to the search path, given that bids do not concern already allocated network resources.
Moreover, two nodes are bound by an edge when the corresponding SPs' bids compete for the same resources. In the algorithm, $f^*$ denotes the best solution found and is regularly updated as better bids are found during the search. Revenue from winning bids on the search path is denoted by $g$. As nodes (selected bids) are removed from $\mathcal{G}$ down a search path, their edges are also removed. Similarly, as nodes are reinserted into $\mathcal{G}$ when backtracking, edges are also reinserted. To prune across independent components of $\mathcal{G}$, $\lambda_{min}$ is used to denote the minimum revenue that the InP expects from a given SP's bid \cite{Sandholm2005}. Furthermore, it evaluates whether a search should be pruned based on the unallocated items, estimated by the upper and lower revenue bounds.

\subsubsection{{Scenario 1}}
For this scenario, we first upgrade CABOB with a fairness mechanism. To do so, the InP artificially inflates losing SPs' bids at the end of each time period $t$, using a weight parameter $\Delta_s^t$. The latter is increased after each bid loss. Its value is reinitialized to the smallest value once the SP wins a bid within $\tau_{th}$.  
We also consider a function to gauge the interest sparked by the InP's resources over time. To this end, we note the expected difference between the InP's desired minimum resource price $\lambda_{c,min}^t$ and SPs' bids, a variant of the metric known in the literature as the bid-ask spread. Obviously, the InP's objective is to minimize the bid-ask spread by encouraging competition in the market. It could be argued that the InP can also reduce temporarily its minimum price (ask) to induce market rally, at the expense of a short profit loss, but subsequent gains would be obtained at a later time. For this, the InP needs to keep historical data from bidding rounds.

\subsubsection{{Scenario 2}}
In this scenario, it is critical for the InP to accurately distinguish occurrences of \textit{coopetition} and \textit{collusion} given their similarity in the cooperation process. Indeed, actors in a coopetition strategy work to ensure their stability and viability within the market, whereas actors in a collusion strategy impede market's competition in order to increase their profits at the expense of the auctioneer's and other actors' welfare \cite{Pathak2013}. Here, the InP leverages a simple reputation-based approach, where a score $\vartheta_s = \sum_{t \in \mathcal{T}} \vartheta_s^t$ is associated to each SP, based on the latter's behaviour over time and its impact on the InP's and SPs' profits, such that: 
\begin{equation}
    \vartheta_s^t = \rho_s^t + \sum_{s' \in \mathcal{S}\backslash\{s\}} m_{s,s'}
\end{equation}
where $\rho_s^t$ is the utility score of SP $s$ in time $t$, calculated regarding its own profit and the market value of the resources, and $m_{s,s'}$ is the similarity factor between SPs $s$ and $s'$. $\rho_s^t$ is given by:
\begin{equation}
    \rho_s^t = q^{t'/\mathcal{T'}} \cdot |b_s^t - \lambda_t^{c,min}|_a 
\end{equation}
where $|.|_a$ is the absolute value, $q$ denotes an exponential growth parameter with $\mathcal{T'}$, an observation period, and $t'$ is an incremental factor, increased when the bid $b_s^t$, with a limited utility to SP $s$, persists from a previous observation period, and resets to $0$ otherwise.
Whereas, $m_{s,s'}$ can be written as:
 \begin{equation}
     m_{s,s'} = \sqrt{(b_s^t -b_{s'}^t) + (u_s^t - u_{s'}^t)},\; \forall (s,s') \in \mathcal{S}^2.
 \end{equation} 
 Hence, the InP maintains a similarity matrix of all SPs as follows:
 \begin{equation}
 \textbf{M} = 
 \begin{pmatrix}
  m_{1,1}  & \cdots & m_{1,|\mathcal{S}|} \\
  m_{2,1}  & \cdots & a_{2,|\mathcal{S}|} \\
  \vdots   & \ddots & \vdots  \\
  m_{|\mathcal{S}|,1}  & \cdots & m_{|\mathcal{S}|,|\mathcal{S}|} 
 \end{pmatrix}.    
 \end{equation}
The above information helps the InP evaluate whether an SP's behaviour is more geared towards its own profit or aims at inhibiting competitors.
When the InP is confident about a collusive cooperation between two or more SPs, it attempts to break this partnership by granting access to its resources to only one of the colluding SPs for a given time period. Similarly to Scenario 1, ill-behaved (collusive) SPs are temporarily penalized by artificially deflating their bids' values.



\subsection{From the SPs' Perspective : A Distributed Multi-agent Deep Reinforcement Learning Approach}

From the SPs standpoint, the bidding problem can be modeled as a multi-agent system, where each agent's knowledge and strategic actions are specific to them and dependent on their budget constraints or their perception of the auction environment. We use a multi-agent reinforcement learning approach that can be considered as an extension of Markov Decision Process (MDP), called Markov Game. In this game, $|\mathcal{S}|$ SPs bid for the InP's resources, to satisfy their subscribers' needs. A Markov game is defined by a set of states $\mathcal{X}$ describing the possible status of all bidding agents, a set of actions $\{\mathcal{A}_1,\mathcal{A}_2,\ldots,\mathcal{A}_{|\mathcal{S}|}\}$ with $\mathcal{A}_s$ the action space of SP $s$. 
At each time period $t$, each SP $s$ uses a policy $\pi : \mathcal{X}_s \longmapsto \mathcal{A}_s$ to determine an action $a_s$, where $\mathcal{X}_s$ is the state space of SP $s$. After the execution of $a_s$, SP $s$ transfers to the next state according to the state transition function $\mathcal{X} \times \mathcal{A}_1 \times \ldots \times \mathcal{A}_{|\mathcal{S}|} \longmapsto \phi(\mathcal{X})$ where $\phi(\mathcal{X})$ indicates the set of probability distributions over the state space. Each SP $s$ obtains a reward based on a function of the state, and all the SPs' actions as $R_s^t : \mathcal{X} \times \mathcal{A}_1 \times \ldots \times \mathcal{A}_{|\mathcal{S}|} \longmapsto \mathcal{R}$. 
Each SP $s$ maximizes its own total expected revenue $R_s^{Total}$ such that:
\begin{equation}
 R_s^{Total} = \sum_{t \in \mathcal{T}} \gamma^t \cdot R_s^t, 
\end{equation}
where $\gamma^t \in [0,1]$ denotes a discount factor meta-parameter that underlines the perceived importance of future rewards. Specifically, a factor of 0 will make the agent short-sighted by only considering current rewards, while a factor approaching 1 will make it strive for a long-term high reward. Hereafter, we identify the states, actions and rewards in our auction environment:
\begin{itemize}
  \item \textbf{State}
  Our state design aims at letting any SP optimize its bidding decisions based on its perception of the auction environment. We design the agent state to consist of  
  the QoS requirements $u_s^t$ of its submitted request(s), its bid $b_s^t$, whether the bid of SP $s$ was accepted or not by the InP $y_{s,i}^{t,c}$, $\theta_{s,s'}^t \in \{0,1\}$ indicating whether the agent $s$ decides to cooperate with another agent $s'$ or not, the incurred penalty $\delta_s^t$ and the current losing streak $l_s^t$. 

  Hence, the state for SP $s$ is composed as:
\begin{equation}\label{eq:agentState} 
    \mathcal{X}_s=\{ u_s^t, b_s^t, y_{s,i}^{t,c},\theta_{s,s'}^t,\delta_s^t, l_s^t\}.
\end{equation}
  \item \textbf{Action}
  Every SP $s$ has to take a bidding decision at time $t$ in order to ensure its bid is/remains selected by the InP, for the sake of its subscribers' satisfaction and its own profit. Depending on the environment's state, it must decide whether to adjust or not its next bid, while keeping in mind budget constraints. In a cooperative scenario, it must determine how to bid with another agent. An agent's action $a_s$ can be given by:
\begin{equation}
    a_s(\mathcal{X}_s)= \{b_s^{t-1}+\mu_s^t\} \cup \{\theta_{s,s'}^t\},
\end{equation}
where $\mu_s^t \in \mathbb{Z}:\mu_s^t \in [-b_s^{t-1},(\mathcal{B}_{s}^t-b_s^{t-1})]$.
It must be noted that, even though constrained, our action space remains very large. For computation efficiency purposes, we discretize the action space by defining for each agent a set of legal bids, denoting a range of bidding behaviour from prudent to aggressive. More specifically, each agent is free to bid a portion or the totality of its maximum bid $b_{s,max}^t$. \textcolor{black}{Bid-action space discretization is implemented through equal-width binning \cite{Dougherty1995} with a predetermined number of bins. }
  \item \textbf{Reward}
The agent's reward is formulated as:
\begin{equation}\label{eq:reward}
    R_s^t = (y_{s,i}^{t,c} \cdot (b_{s,max}^t - b_s^t )) - \delta_s^t.
\end{equation}
In this formulation, an agent gains significant profit if it manages to win in the auction by bidding as low as possible. It is worth noting that it is possible for coopeting agents to win in the auction and still incur penalties if a portion of their subscribers are left unserved. In contrast, when an agent takes too much risk by submitting below market bids and as a result loses in the auction, it will be penalized. 
\end{itemize}

\textcolor{black}{A key aspect for the SP, with regards to computational efficiency, consists of its training time. Several works \cite{Shallue2018, Dean2012, Gholami2018, Daniely2014, Bartlett2002,Livni2014,Kearns1990,Boob2018} emphasize the increased scale of parallelism available for neural network training, made possible by hardware advancements. Hence, assuming optimized data parallel systems that spend negligible time synchronizing between processors and leveraging practical training “tricks" (e.g., regularization, over-specification, adequate activation functions, batch size increase, etc.), training time can be reduced and measured in the number of training steps \cite{Shallue2018, Livni2014}. For instance, \cite{Shallue2018} showed the correlation between reduced training time and increasing batch size without degradation of the solution quality.}

\textcolor{black}{\subsection{Bidding analysis}
The objective of the auction framework presented above is two-fold : (1) assign resources to bidding who value them the most while (2) ensuring as well profitable revenues for the InP, particularly in cases where SPs have heterogeneous budgets and in the presence of collusion. While truthfulness is the dominant strategy in Vickrey auctions \cite{Vickrey1961}, a noted shortcoming consist of its inability to guarantee auctioneer revenues \cite{Ausubel2004}. Our framework counters by allowing the InP to artificially introduce competition in both scenarios, respectively applying a fairness and a collusion detection/punishment mechanism in addition to enabling coopetition for poor agents.  
}
\textcolor{black}{
Consider $b_{s,max}^t$ as the true value of resources (which are indivisible) for an SP $s$. The payoff received is given in eq.(\ref{eq:reward}). Allow $\max\limits_{s'\neq s} b_{s'}^t $ as the highest bidder $s'$ among competing SPs other than SP $s$. Recall the InP allocation policy, which consists of always selecting the highest bids. Furthermore, an SP $s$ cannot bid above its budget. 
These are the following outcomes following a bid by SP $s$ for a time period $t$:
\begin{equation}\label{eq:true1}
    \max\limits_{s'\neq s} b_{s'}^t < b_{s,max}^t
\end{equation}
\begin{equation}\label{eq:true2}
    \max\limits_{s'\neq s} b_{s'}^t < b_{s}^t~|~b_{s}^t < b_{s,max}^t 
\end{equation}
\begin{equation}\label{eq:true3}
    \max\limits_{s'\neq s} b_{s'}^t > b_{s,max}^t
\end{equation}
\begin{equation}\label{eq:true4}
    \max\limits_{s'\neq s} b_{s'}^t  > b_{s}^t~|~b_{s}^t < b_{s,max}^t
\end{equation}
\begin{equation}\label{eq:true5}
    b_{s}^t < \max\limits_{s'\neq s} b_{s'}^t  > b_{s,max}^t
\end{equation}
In eq. (\ref{eq:true1}) and (\ref{eq:true2}), SP $s$ wins both with a truthful bid, and underbids with a positive payoff for an underbid compared to a zero penalty paid for truthful bidding. In eq. (\ref{eq:true3}) and (\ref{eq:true4}), the SP loses the bid and incurs a penalty being paid regardless of the employed bidding strategy. The only winning strategy in eq. (\ref{eq:true5}) is truthful bidding with zero payoff while other bidding strategies induce a penalty. Hence, it can be seen that truthful bidding is the dominant strategy in every case except in eq. (\ref{eq:true2}) as underbidding is the dominant strategy.
}

\textcolor{black}{In our framework, the weak dominance of truthful bidding is designed as an incentive to prolong market competitiveness. It allows for poor SP $s$ to remain within the market given it can adequately adapt its bidding strategy. For richer SPs, underbidding is a risky strategy as it gives poor SPs an opportunity to win the auction. Furthermore, in the scenario where a fairness mechanism is applied, an underbidding SP would probably lose in favour of a poor SP with an artificially inflated bid. Collusive SPs attempt to bid truthfully in the short term, hoping to drive poor SPs out of the auction, and aiming to underbid in the long term. We counteract this set of strategies by detecting collusive SPs at first. Once this is done, we create a cost asymmetry between the colluders by granting access to resources to only one SP, which constitutes an effective barrier to collusion \cite{Ivaldi2003}.}

\section{Experimental evaluation}
For our evaluations, we leveraged the Open AI Gym toolkit to implement a custom auction simulation environment, where SPs bid to gain access to resources to satisfy their subscribers' demands. As mentioned in Section \ref{InpSolutions}, the InP runs a modified version of CABOB, and activates in certain scenarios a fairness mechanism favouring underserved SPs within the market. \textcolor{black}{We leave the case of multiple InPs and its analysis for future work.}
We consider 8 independent bidding agents (i.e. SPs) who apply specific bidding policies, learned after 1000 episodes of training. Auction environment parameters are summarized in Table \ref{sim_settings}. 

Given our action space, we evaluate and compare the following bidding algorithms in similar simulation settings:
\begin{itemize}
    \item \textit{Incremental}: An agent using this algorithm applies a simple conservative bidding strategy. When it wins in a bidding round, it linearly decreases its bid. In contrast, when the agent suffers an auction loss, it exponentially increases its bid in the following round. \textcolor{black}{ Incremental is designed to mimic the behaviour of two algorithms. Similarly to AWESOME \cite{Conitzer2007}, it assumes rival agents behavior at different stages of the game and tries to maintain hypotheses about an agent's behaviour. Like GIGA-WoLF \cite{Bowling2005}, it also uses an adaptive step that makes it more or less aggressive in changing its bidding strategy.}
    \item \textit{Random}: In this benchmark strategy, an agent randomly selects an action within the legal action set.
    \item \textit{DQN (Deep Q-learning):} DQN is a variant of the well known Q-learning technique using a deep neural network for stable learning \cite{Mnih2015}. It uses the experience replay technique, where random samples of previously stored experiences are taken into account for future learning and action selection. 
    \item \textit{DDQN (Double DQN):} DDQN aims at reducing the overestimation of Q values, encountered in DQN \cite{Hasselt2010}. Hence, it allows faster training and a more stable learning. It leverages two policies, one for value evaluation and another for future action decisions. 
\end{itemize}

\begin{table}[t]
    \centering
       \caption{Sim. Parameters}
    \begin{tabular}{|l|c|}
 \hline
\multicolumn{2}{|c|}{\textbf{Auction settings}} \\ \hline
Number of bidding rounds                 & 10\\ \hline
Number of SPs $\mathcal{S}$                & 8\\ \hline
Number of InPs                 & 1\\ \hline
Fairness threshold $\tau_{th}$ & 3 \\ \hline
\textcolor{black}{Number of action space discretization bins} & \textcolor{black}{10} \\ \hline
\multicolumn{2}{|c|}{\textbf{InP settings}} \\ \hline
Number of subcarriers                & 10 \\ \hline
\multicolumn{2}{|c|}{\textbf{SP settings}} \\ \hline
Number of subcarriers required to satisfy QoS               & 4\\ \hline
Penalty after bid rejection               & 25\\ \hline
Number of subscribers & 3\\ \hline
Number of bidding behaviours & 6\\ \hline
\multicolumn{2}{|c|}{\textbf{Hyperparameters}} \\ \hline
Learning rate                 & 0.001\\ \hline
Memory size $\mathcal{S}$                & 10000\\ \hline
Batch size                 & 128\\ \hline
Probability $\epsilon$                 & 1 - 0.01\\ \hline
Probability $\epsilon$ decay                & 1 - 0.01\\ \hline
Number of episodes & 4000 \\ \hline
Discount factor $\gamma$ & 0.95 \\ \hline
    \end{tabular}
 
    \label{sim_settings}
\end{table}

\begin{table}[t]
    \centering
      \caption{Budget Distribution}
    \begin{tabular}{|c|c|c|}
 \hline
\cellcolor[HTML]{9B9B9B} & \textbf{Number of agents} & {\textbf{Budget range}} \\ 
\hline
\multicolumn{3}{|c|}{{\color[HTML]{333333} Full competition}} \\ \hline
\multicolumn{1}{|l|}{Competing agents} & 8 & \multicolumn{1}{l|}{300 - 1000} \\ \hline
\multicolumn{1}{|l|}{Colluding agents} & 0 & - \\ \hline
\multicolumn{1}{|l|}{Coopeting agents} & 0 & - \\ \hline
\multicolumn{3}{|c|}{Collusion} \\ \hline
Competing agents & 6 & 300 - 1000 \\ \hline
Colluding agents & 2 & 1000 \\ \hline
Coopeting agents & 0 & - \\ \hline
\multicolumn{3}{|c|}{Collusion and coopetition} \\ \hline
Competing agents & 2 & 500 - 1000 \\ \hline
Competing agents & 2 & 300 - 500 \\ \hline
Colluding agents & 2 & 1000 \\ \hline
Coopeting agents & 2 & 300 \\ \hline
    \end{tabular}
  
    \label{variables3}
\end{table}

\textcolor{black}{The baseline approaches presented above aim to highlight specific insights on the behaviour of deep learning-driven agents in dynamic market interactions. We omit multi-agent learning algorithms such as Fictitious Play \cite{Brown1951}, Bully \cite{Littman2001}, AWESOME \cite{Conitzer2007}, Meta \cite{Powers2005}, Minimax-Q \cite{Littman1994}, Nash-Q \cite{Hu2003}, Correlated-Q \cite{Greenwald2003}, GiGA-WoLf \cite{Bowling2005}, $RV_{\sigma(t)}$ \cite{Banerjee2006}, and GSA \cite{Spall2005}, given they have all been shown to be less performing and stable than Q-learning \cite{Zawadzki2014}. Evolutionary-based algorithms are also omitted due to their drawbacks and for fairness purposes. Indeed, evolutionary algorithms cannot guarantee optimality. In addition, the associated solution's quality depends highly on the initialization setting, and it deteriorates as the problem size increases.}

Based on the scenarios of section \ref{auction_desc}, three types of auction environments are assumed: 1) Full competition, 2) collusion with fairness and 3) collusion and coopetition. In the first, we investigate the market dynamics in which agents are seeking access to the InP's resources without any form of cooperation. In the second, some agents may coordinate their bids to increase their mutual profits. Specifically, we consider 2 colluding and 6 competing agents within the marketplace. In the last case, we assume 2 colluding, 2 coopeting and 4 competing agents within the marketplace. Moreover, we qualify the agents with a starting budget within the range 800 to 1000 as rich agents, within 500 to 800 as middle budget agents, while agents with budgets below 300 are called poor agents. The environment parameters are summarized in Table \ref{variables3}.

\begin{figure}[t]
    \centering
    \includegraphics [width = 0.99 \columnwidth]{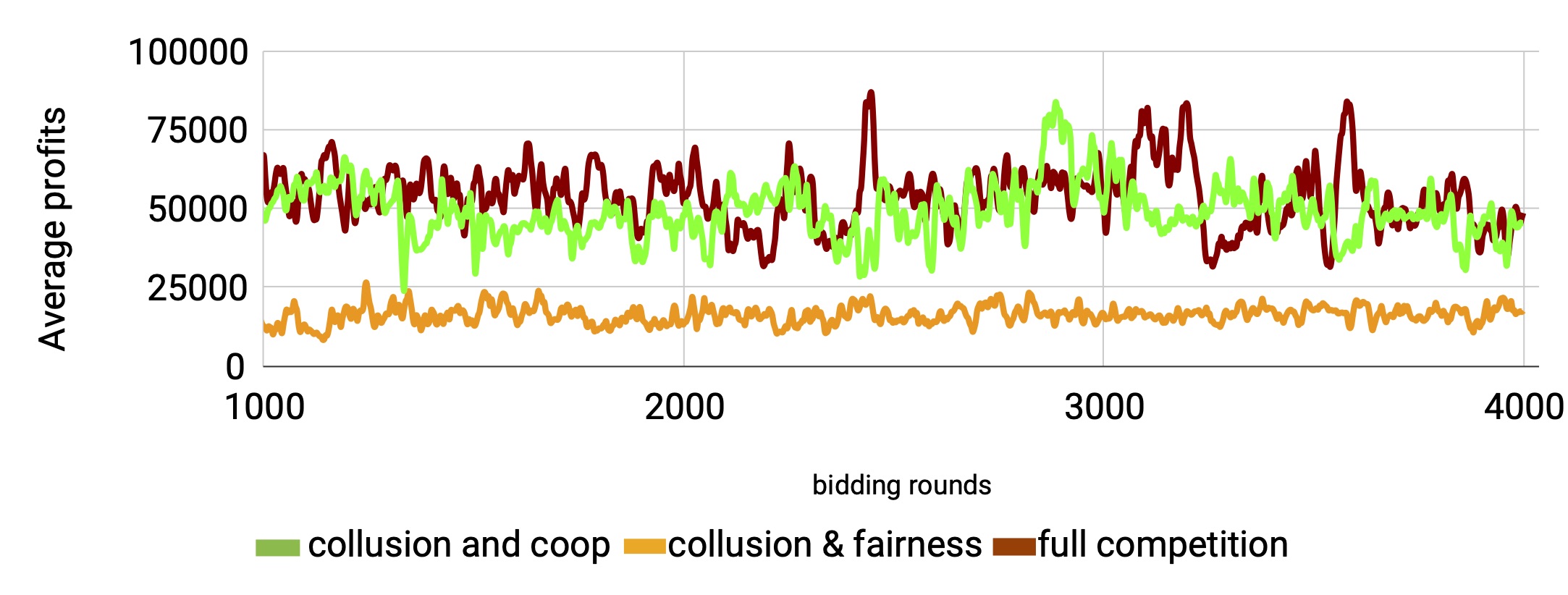}
    \caption{Average InP profits in different auction environments.}
    \label{fig:AvgInP}
\end{figure}

\subsection{From the InP perspective}
First, we investigate the InP's average profits under different auction environments as shown in Fig. \ref{fig:AvgInP}. According to simulations, the most profitable auction environment is full competition. In this case, DRL-aided SPs rely only on themselves to win in the auction market. We observe a general tendency where SPs typically react by bidding aggressively after an auction loss, thus leading to a profit increase for the InP. Similarly, when an agent achieves sustained auction wins, it seems to regularly lower its bids, in an attempt to increase its own profits. Such behaviour is highlighted, for instance in Fig. \ref{fig:profits_sp_comp} where a DDQN-driven SP is able to adapt its bidding behaviour quickly after consecutive auction losses.
\begin{figure}[t]
\centering
    \includegraphics [width=0.99 \columnwidth]{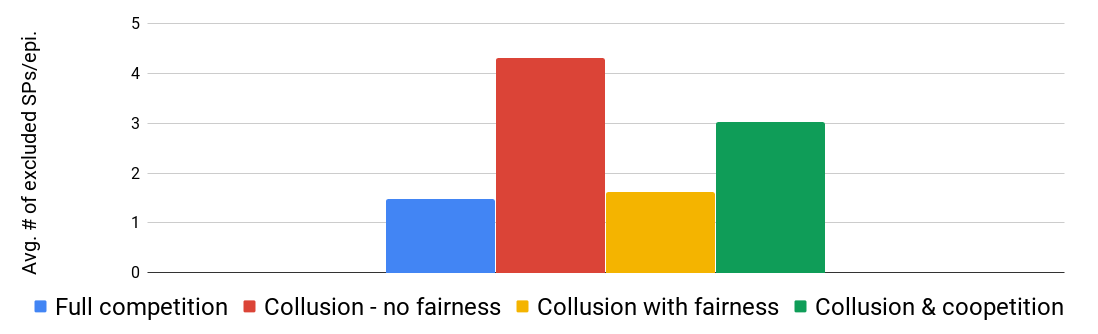}
    \caption{Average number of excluded SPs per episode.}
    \label{fig:excl.}
\end{figure}
With similar profitability, the collusion and coopetition auction setting can arguably be viewed as a form of full competition. Indeed, the InP counter-attacks collusion by empowering poor agents to jointly bid to access the resources and avoid quitting the market. As a consequence, colluding agents are forced to over-bid and enrich the InP. However, colluding remains able to profit rich agents at the expense of poor ones. In an attempt to remedy to this situation, we proposed a fairness mechanism, where the InP allow SPs to access its network resources within a certain threshold ($\tau_{th} = 3$). This setting is shown to be the least profitable, since enforced fairness comes at the expense of its profits.

\begin{figure}[t]
\centering
    \includegraphics [width=0.99 \columnwidth]{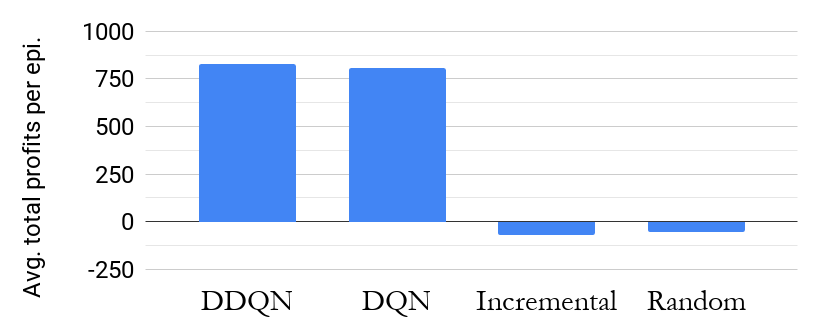}
    \caption{Average total profits per episode under each algorithm.}
    \label{fig:algo_comp}
\end{figure}
\begin{figure}[t]
\centering
    \includegraphics [width=0.99 \columnwidth]{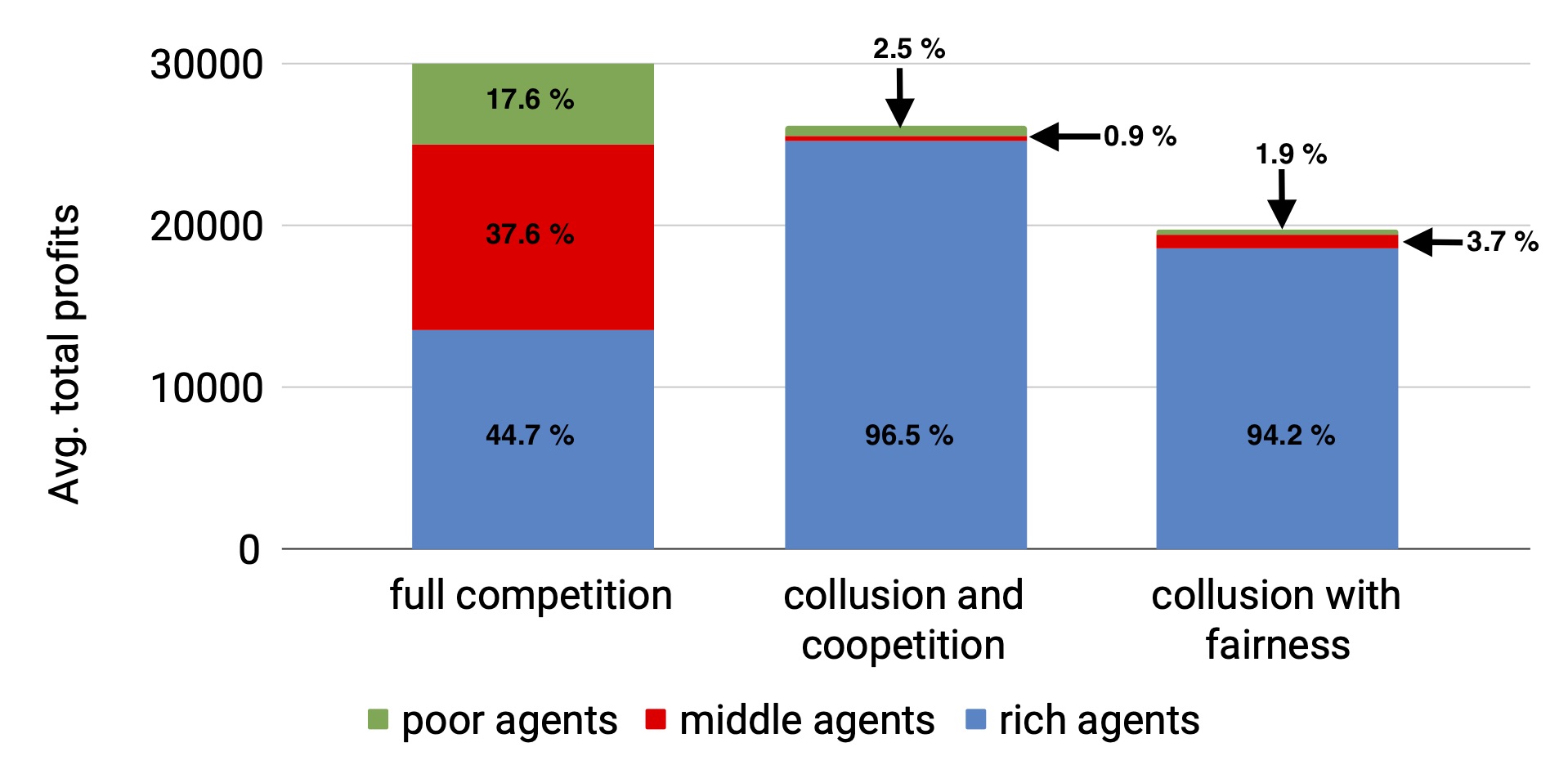}
    \caption{Average total SP profits under different auction settings.}
    \label{fig:profits_sp_budget}
\end{figure}

In Fig. \ref{fig:excl.}, we compare the average number of SPs forced out of market. The highest number of exclusions is in the collusive environment. Indeed, since collusive agents have high budgets, they are able to consistently bid over market value, leading into sustained auction losses to poor agents. Both full competition and the collusion with fairness settings have the best ability to keep poor SPs in the market. Whereas, collusion and coopetition environment presents a non-negligible number of exclusions. This can be due to the amount of time it takes coopeting agents to adjust their bids and beat colluding agents.    
To be noted that coopetition is alleviated when agents' budgets become above 700 (middle budget agent) due to accumulated profits.

\subsection{From the SPs perspective}
\begin{figure}[ht]
\centering
    \includegraphics [width=0.99 \columnwidth]{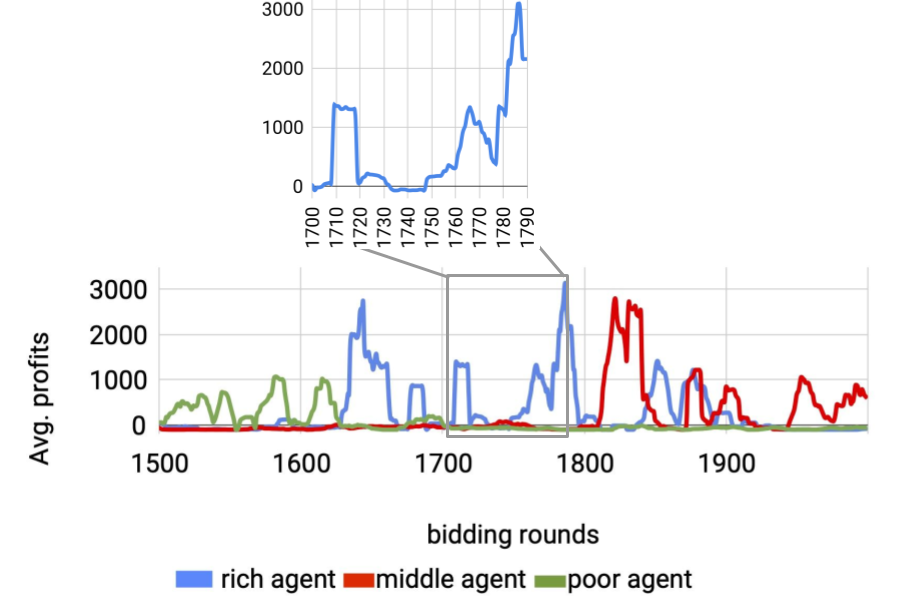}
    \caption{Average SP profits under full competition auction environment.}
    \label{fig:profits_sp_comp}
\end{figure}

In Fig. \ref{fig:algo_comp}, DDQN, DQN, incremental and random algorithms are compared, in terms of average profits gained by an SP. Both incremental and random algorithms present the worst profits. Whereas, DQN and DDQN achieve very high performances. Indeed, these DRL-based approaches enable SPs to smartly adjust their bidding strategies according to market conditions. 
Indeed, as it can be seen in Fig. \ref{fig:profits_sp_comp}, the DDQN agent is able to recover after a series of auction losses.  
In the remaining simulations, only DDQN agents will be considered, for their fast convergence and better performances, compared to the other approaches. 

In Fig. \ref{fig:profits_sp_budget}, we illustrate the SPs' profits as a function of auction environments, and for different SPs' profiles. We assume that rich agents are initiated with a budget of 1000, middle agents with a budget of 500 and finally poor agents with a budget of 300. Meanwhile, the same number of subscribers and QoS requirements are given to them. Results show that the full competition environment is the most beneficial to both middle and poor agents. Interestingly, this setting showcases that middle agents and rich agents are able to compete with comparable profit margins despite their budget differences, thus indicating a potential for greater profits if agents are able to adapt more precisely their bidding behaviour. This can be seen in Fig.\ref{fig:profits_sp_comp} where each agent is able to secure wins and most importantly offsets auction losses with profits margins mostly above 0 during the auction process. Moreover, poor agents' strategy seems to require short aggressive bidding bursts in order to win, with a downside of limited bidding power afterwards. Finally, it is shown that the fairness mechanism, while improves middle agents' performance, has a very small impact on the rich agents' profits. Indeed, SPs' intelligence allows them to detect the fairness mechanism and bypass it, by smartly adjusting their bids. Consequently, colluding agents still get most profits.

\begin{figure}[t]
\centering
    \includegraphics [width=0.99\columnwidth]{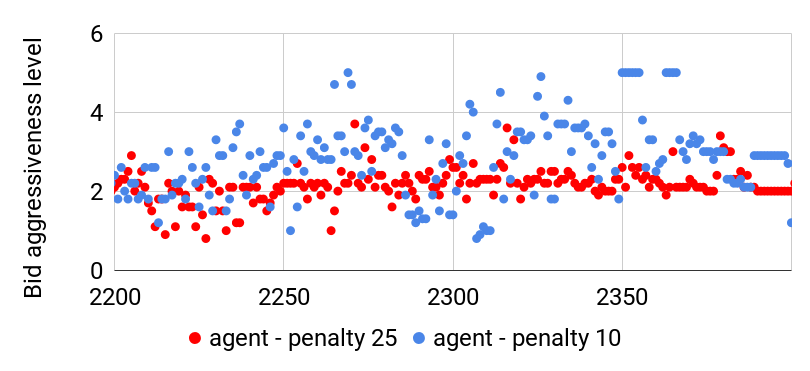}
    \caption{Impact of penalty function on agent behaviour.}
    \label{fig:agent_pen}
\end{figure}

The impact of the penalty's value is investigated, by illustrating agents' bidding behaviours in Fig.\ref{fig:agent_pen} where the vertical axis denotes the level of bid aggressiveness adopted by an agent. 
Here, we compare two agents operating under a different penalty coefficients. As such, we observe a trend where the agent with a lower penalty coefficient seems to adopt a risky bidding behaviour by bidding less aggressively perhaps to conserve its limited budget achieving occasionally high value benefits. In contrast, the agent with a higher penalty coefficient seems more focused on offsetting auction losses.

A critical benefit of the proposed approach is the potential for self management (including self negotiation and self organization) as highlighted in \cite{Marinescu2014} when SPs are allowed automatically switch between InPs whenever necessary for orchestration or profitability reasons. Hence, it can be argued that DRL-based agents could self manage VNFs by switching serving InP once the market become detrimental to them (i.e profit loss, presence of collusion, etc.). Sometimes however, there are obstacles toward InP switching. For instance in mobile networks where the number of InPs to switch is limited by cellular coverage. Further, from the simulations above, poor agents in auction settings where collusion is present are the most likely to regularly switch InP in order to ensure their end-users QoS requirements are met. This entails however that the agent must re-train as they are put in a new environment, which may hinder their performance. 

\section{Discussion}

\subsection{Context}
\textcolor{black}{It has become common belief that AI will become a focal point, not only for network management, but also for next generation applications that are expected to generate and consume exponential quantities of data, and requiring run-time processing before and after transmission, and in some cases, on their way to the end-user \cite{Zhani2019}. This comes as next generation networks are increasingly being described to be independent and decentralized systems on which decisions are taken at different granular levels \cite{Zhani2019, Huang2019}, and based on numerous requirements. Recently, the ITU-T Focus Group Technologies for Network 2030 \cite{network2030, network2030_v2} have pushed forward the idea of “Manynets", on which they state: \textit{[...] Quite likely there will not be just one, but many public Internets. New technologies further widen  the  constraints  for  transmitting  packets  through  the  utilization  of  infrastructure-based  wireless, wireless mesh, satellite, fixed line technologies (such as fibre optics), all of which must be accompanied by the fundamental packet transfer solution, while adhering to the underlying ownership relations when traversing those different networks. As a consequence, the end-to-end realization of services across those many internet environments need strong consideration for Network2030 and is an increasing departure from the structures of networks as we see today.}}

\textcolor{black}{
An example of such requirements, notably for AI-driven applications, is multi-flow synchronization. The latter ensures that generated data from different sources arrive at the destination within a specific interval of time or even at a particular point of time \cite{Zhani2019}. For example, paradigms such as federated learning \cite{Konevcny2016}, where models are trained on network resources located on edge using local sample patterns and sent to centralized entity to build a shared global model, are leveraged to speed up and improve the distributed learning process \cite{Huang2019}, strengthen security and guarantee privacy. 
}

\textcolor{black}{
Several observations motivate the use of auction-based solutions proposed in this work. Firstly, as network resources are distributed on independent domains, it is obvious that a large-scale autonomous cooperation/negotiation mechanism must be in place to ensure access to adequate network resources. Second, with NFV as an enabling technology, the cooperation mechanism must also be able to embed dynamic service requirements in order to achieve multi-domain orchestration. In this regard, we argue that auction and market dynamics are a flexible signal mechanism to identify adequate resource locations, allowing for instance network characteristics (penalty to compensate for low bandwidth, high latency, etc.) to be easily embedded within a bid, aiming to jointly solve the profitability and performance conundrum.
}

\textcolor{black}{
However, given that bidding agents will most likely be powered through AI-enabled algorithms, it is essential to investigate the impact of such agents within the market, and effective methods to counter them. In fact, the Organisation for Economic Co-operation and Development (OECD) reports that there is a particular concern for AI-enabled algorithms to become \textit{a  facilitating factor for collusion and may enable new forms of co-ordination that were not observed or even possible before. This is referred to as “algorithmic collusion”} \cite{oecd}. 
}

\subsection{Architectural implications for Future Networks}
\textcolor{black}{
Despite strong auction performance dominance for DRL-based agents compared to other agents after training, two main concerns still need to be addressed. The first concern is to investigate into where the training process will take place. In this regard, the ITU-T Focus Group on Machine Learning for Future Networks has advocated for a sandbox domain, which is an internal operator where machine learning (ML) models can be trained, verified and their effects on the network studied \cite{Itusandbox}. The second concern relates to the amount of communication between coopeting/colluding agents that may induce further congestion in future networks. 
}
\subsection{Multiple InPs}
\textcolor{black}{While we considered scenarios with a single InP, we hypothesize that considering multiple InPs scenarios may induce several changes to both InPs and SPs strategies. For instance, shifting the level of aggressiveness an SP would display to obtain networking resources given that alternative choices also exist. Naturally, this influences whether and when (in particular, poor) SPs exit the market, as bidding conditions may change favourably or unfavourably. A direct consequence would be that InPs become in a weaker position, especially with the presence of colluding SPs who would in the long run endanger InPs' profits margin. It puts also an increased pressure on the InPs as they must adopt more generous fairness mechanisms to retain self-interested SPs, while severely penalizing colluding SPs. Consequently, this would come at the expense of slightly reduced profits, but still better than a full collusive market. On the other hand, market switching in the context of resource allocation may not be all benefits for SPs. In addition to restarting the process of market price discovery, the possible markets/InPs a given SP can negotiate with may be constrained by its end-user requirements and the incurred penalty's value, assuming it fails to serve its users.}

\section{Conclusion}
As one of the major goals of upcoming 5G networks, the need for end-to-end service provisioning has rendered urgent new profitable cooperation and service orchestration mechanisms particularly for multi-domain networks. We thus proposed a market driven in which SPs and InP interact to exchange and orchestrate resources while keeping in mind stringent QoS requirements. Through realistic market scenarios, we analyzed the behaviour of DRL-based agents. Our simulations results have shown DDQN and DQN - driven agents to perform generally well in dynamic network environments, although profitability has been shown to be constrained by budget limitations. From the InP perspective, a competitive auction environment has been shown to be the most profitable. However, in cases where collusion between SPs is present, tradeoffs must be made between ensuring market fairness and harming its profits. 

\ifCLASSOPTIONcaptionsoff
  \newpage
\fi

\bibliographystyle{IEEEtran}
\small
\bibliography{TNSM}

\end{document}